\documentclass[aps,twocolumn,pra,%
preprintnumbers,amsmath,amssymb,superscriptaddress]{revtex4}
\usepackage{graphicx}
\usepackage{color}
\newcommand{\bea}{\begin{eqnarray}}
\newcommand{\eea}{\end{eqnarray}}

\newcommand{\tr}{\mathrm{Tr}}

\newcommand{\mc}[1]{\mathcal{#1}}
\newcommand{\mb}[1]{\mathbb{#1}}
\newcommand{\Lbar}[1]{\overline{#1}}

\begin{document}

\preprint{OU-HET-857}
\preprint{RIKEN-STAMP-5}
\preprint{YITP-15-31}

\title{Exact Path Integral for 3D Quantum Gravity}

\date{\today}

\author{Norihiro Iizuka}\email[]{iizuka@phys.sci.osaka-u.ac.jp} 
\affiliation{{\it Department of Physics, Osaka University, Toyonaka, Osaka 560-0043, JAPAN}}

\author{Akinori Tanaka}
\email[]{akinori.tanaka@riken.jp}
\affiliation{{\it Department of Physics, Osaka University, Toyonaka, Osaka 560-0043, JAPAN}}
\affiliation{{\it Interdisciplinary Theoretical Science Research Group, 
RIKEN, Wako 351-0198, JAPAN}}

\author{Seiji Terashima}
\email[]{terasima@yukawa.kyoto-u.ac.jp}
\affiliation{{\it Yukawa Institute for Theoretical Physics, 
Kyoto University, Kyoto 606-8502, JAPAN}}

\begin{abstract}
Three dimensional Euclidean pure gravity with a negative cosmological constant can be formulated in terms of the Chern-Simons theory, classically. This theory can be written in a supersymmetric way by introducing auxiliary gauginos and scalars. We calculate the exact partition function of this Chern-Simons theory by using the localization technique. Thus, we obtain the quantum gravity partition function, assuming that it can be obtained non-perturbatively by summing over partition functions of the Chern-Simons theory on topologically different manifolds. The resultant partition function is modular invariant, and in the case in which the central charge is expected to be $24$, it is the $J$-function, predicted by Witten.  
\end{abstract}

\maketitle



\noindent

\section{Introduction}\label{sec:intro}
Non-perturbative calculation for pure gravity in three dimensions is a very interesting problem.  
It contains no local degrees of freedom, 
but it contains black hole solutions \cite{Banados:1992wn}. 
In this paper, using the Chern-Simons formulation \cite{Achucarro:1987vz,Witten:1988hc} and recently developed localization techniques, we calculate 
the partition function of the three-dimensional pure gravity in asymptotic Anti-de Sitter space-time,   
\bea
\label{GRpathintegral}
Z_{gravity} = \int {\cal{D}}g_{\mu\nu} e^{- S_{gravity}} 
\eea
where, three dimensional pure gravity action is 
\bea
S_{gravity} =\frac{1}{16 \pi G_N} \int d^3x \sqrt{g} \left( R - 2  \Lambda \right)   
+ S_{GH} + S_c \,. 
\label{gravityaction}
\eea
Here $\Lambda = - 1/\ell^2$, and $\ell$ is AdS scale, and 
$S_{GH} = \frac{1}{8 \pi G_N} \int d^2x \sqrt{h} K$ is Gibbons-Hawking boundary term with $K$ extrinsic curvature, and $h$ the boundary metric, 
and $S_c = - \frac{1}{8 \pi G_N} \int d^2x \sqrt{h}$ is an appropriate gravity counter-term.  
We focus on negative cosmological constant case since asymptotic AdS 
gives possibly well-defined boundary condition for quantum gravity \cite{Maldacena:1997re}.  

Our strategy is as follows;
First, we utilize the following well known relationship between the Chern-Simons actions and the 1st order gravity action  \cite{Achucarro:1987vz,Witten:1988hc}:
\bea
&&  \int_{\cal{M}}  d^3 x \ e \Big(R(e,\omega) + \frac{2}{\ell^2} \Big) + 
\int_{\partial {\cal{M}}}  e^a \omega^a \notag \\ 
&& \qquad = i \ell \Big( S_{CS}[\mc{A}] -  S_{CS}[\Lbar{\mc{A}}] \Big)
\,,
\label{key} 
\eea
where we introduce the $SL(2,\mb{C})$ gauge valued Chern-Simons action 
\bea
S_{CS}[\mc{A}]
:=
\int_{\mc{M}} \text{Tr} \Big( \mc{A}d \mc{A} + \frac{2}{3} \mc{A}^3 \Big) \,,\,\,\,\,\,\,\, \qquad \\
\label{garelation}
\left( \omega_\mu^a + \frac{i}{\ell} e_\mu^a \right) \frac{i}{2} \sigma_a  dx^\mu = \mc{A} \,, 
\left( \omega^a_\mu - \frac{i}{\ell} e^a_\mu \right)  \frac{i}{2} \sigma_a  dx^\mu = \bar{\mc{A}} \,, 
\eea
where $\sigma_a$ is Pauli matrix. 
Then, defining $k \equiv \ell/4 G_N$ \cite{Carlip:1998qw}, 
we can write gravitational action (\ref{gravityaction}) as  
\bea
\label{gravity_gauge}
S_{gravity}  = \frac{i k}{4 \pi} S_{CS}[\mc{A}] - \frac{i k}{4 \pi} S_{CS}[\bar{\mc{A}}] 
\equiv S_{gauge} \,.   
\eea
Here, we use an important fact: on-shell value of $\frac{1}{16 \pi G_N} \int_{\partial \mc{M}} e^a \omega^a$ 
term in \eqref{key} gives a contribution corresponding to on-shell value of $S_{GH}+S_{c}$ in \eqref{gravityaction}.

From the work of Brown-Henneaux \cite{Brown:1986nw}, 
the central charge $c$ of this theory is given by $c = 3 \ell /2 G_N$, which is equal to 
$c = 6 k$  in our convention.  Quantization condition for $k$, for $SL(2, \mb{C})$ gauge group, is 
$k \in \mb{Z}$, and this gives $c/ 6 \in \mb{Z} $. However since 
the diffeomorphism of Euclidian $AdS_3$ is $SO(3,1)$, 
 precisely speaking, we have to conduct the Chern-Simons theory calculation for 
 the gauge group $SO(3,1)$. 
 In that case, the allowed central charge is $c/24 \in \mb{Z}$ \cite{Witten:2007kt}. 
Due to the fact that the localization of the Chern-Simons theory for $SU(2)$ is well-studied, 
we use the notation of $SL(2, \mb{C})$,  a complexification of $SU(2)$, for the Chern-Simons theory gauge group 
in \eqref{key} for simplicity. This is justified since 
$SO(3,1)$ and $SL(2, \mb{C})$  are locally identical, and 
global difference between them cannot be seen 
at the level of classical action. 
Rather it should give different quantization condition for $k$ as 
$k/4 \in \mb{Z}$. 

Second, to conduct path-integral for gravity, 
we define the measure for quantum gravity in terms of the Chern-Simons theory 
such that 
(\ref{GRpathintegral}) can be calculated in terms of 
the Chern-Simons theory path-integral. 
Note that the metric information $g_{\mu\nu}$ 
in gravity theory is encoded in the gauge boson $\mc{A}_\mu$ in the Chern-Simons theory 
through the relation eq.~(\ref{garelation}). 
The Chern-Simons theory lives 
on three-dimensional Euclidian space, 
which we call $\cal{M}$. 
{Since the Chern-Simons theory is topological, 
it depends on only 
the {\it topology} 
of $\cal{M}$ and 
the boundary metric of $\cal{M}$. These must be determined 
from the corresponding space-time in gravity.  
Since we solve the quantum gravity in asymptotic AdS boundary condition and 
asymptotic Euclidean AdS is parametrised by torus, whose complex moduli is $\tau$, 
$\cal{M}$ must have a boundary torus with moduli 
$\tau$.} 

Note that different space-time topology in gravity side should be  
mapped into different topology for $\cal{M}$ in the Chern-Simons side.     
While in the gauge theory, one regards the Chern-Simons theory 
on different topology $\cal{M}$ as a different theory. 
{Therefore for the correspondence to work, 
we decompose the metric path-integral into each 
sector distinguished by the topology, and then 
we will sum over different topology, where all of them 
should have a boundary torus. 
Furthermore, in the Chern-Simons theory, boundary conditions (b.c.), related by the 
modular transformation for $\tau$, are regarded as giving different theory. On the other hand, 
in gravity, radial rescaling changes the size of torus, and the size of torus cannot be a parameter for the boundary. 
Therefore we also need to sum over different b.c. for $\mc{A}$, where all of the b.c. for $\tau$ are related by the modular transformation.  
We schematically write this decomposition as 
${\cal{D}}g_{\mu \nu} = \sum 
 {\cal{D}}g^{sector}_{\mu \nu} $, 
and then we replace each measure ${\cal{D}}g^{sector}_{\mu \nu}$ by the Chern-Simons theory 
path-integral ${\cal{D}} \mc{A} {\cal{D}} \bar{\mc{A}}$ where $\mc{A}$ lives on $\cal{M}$ with corresponding topology. 
Here we understand that $ {\cal{D}}g^{sector}_{\mu\nu}$ is metric integration for {\it a given} bulk topology and {\it a given} boundary torus. 
}

To understand what is the appropriate summation, 
note that all of the known classical solutions of pure gravity Einstein's equation are AdS space-time and BTZ black holes. 
All of them have the topology $D^2 \times S^1$, which is the same as solid torus. 
Therefore we assume that topology which contributes to the quantum gravity is only the solid torus,  and nothing else.  
This will be justified after the application of the localization technique 
because only the saddle points contribute there, and they satisfy the equation 
of motion for the Chern-Simons theory, which is ``equivalent'' to Einstein's equation. 
Furthermore, all of the classical solutions admit asymptotic AdS space-time, parametrised by torus complex moduli 
$\tau$, and they are related by $SL(2, \mb{Z})$ transformation \footnote{For examples, 
in zero angular momentum case, thermal AdS and zero-angular BTZ black hole are related by S-transformation of $SL(2,\mb{Z})$, $\tau \to - 1/\tau$.}.  
Therefore, the topologically distinct sectors are labeled by specifying
which circle on the boundary is contractable. 
Thus, we need to take summation over the `cosets of $SL(2 ,\mb{Z} )$'. 
Here, cosets of $SL(2, \mb{Z})$ summation is parametrised by $\tau \to (a\tau + b)/(c \tau + d)$ satisfying $ad - bc = 1$, 
where, $(-a,-b,-c,-d) \approx (a,b,c,d)$ makes $c \ge 0$ and $(c,d)_{GCD} = 1$ so that $a$, $b$, are uniquely 
determined up to $(a,b) \approx (a + c, b + d)$ to satisfy $a d - b c = 1$.  
The last ambiguity does not matter since the expression for partition function is written in terms of 
summation over cosets of $SL(2, \mb{Z})$ about variable $exp\left( 2 \pi i \frac{a \tau + b}{c \tau + d}\right)$, and 
the ambiguity $(a,b) \approx (a + c, b + d)$ does not change 
this variable. 

Physically parameter $a, b, c, d$ specifies the choice of a bulk contractable circle 
as a linear combination of $t_E$ and $\phi$ circle. 
Summation over $c$ and $d$ is simply summing over all the choice of classical solutions, which are specified by 
how to choose a bulk contractable circle. 
As a result, final answer after summing over topology is modular invariant.   

In summary, topology summation 
is defined naturally as cosets of $SL(2 ,\mb{Z} )$, 
$ \sum_{c \ge 0, (c,d)=1}$. 
{
In fact, Farey tail story \cite{Dijkgraaf:2000fq,Maloney:2007ud, Manschot:2007zb, Manschot:2007ha} takes exactly the same approach for the  gravity path-integral 
measure, 
where Elliptic genus evaluated in the dual field theory is decomposed as summation over cosets of $SL(2 ,\mb{Z} )$ by Rademacher expansion. 
}

Third, 
we assume that we can regard $\mc{A}$ and $\bar{\mc{A}}$ are independent holomorphic variables. 
Therefore we expect that $\mc{A}$ and $\bar{\mc{A}}$ can have different topology, {\it i.e.,} 
topology of solid torus with different choice of $(c, d)$, a different choice of contractable circles. 
The decomposition of the Chern-Simons theory in terms of $\mc{A}$ and $\bar{\mc{A}}$ in (\ref{gravity_gauge}) motivates  us to assume the holomorphic factorization \cite{Witten:2007kt} as Witten. 
All of these considerations lead us to the relationship  
\bea
\sum {\cal{D}}g^{sector}_{\mu \nu}  = \left(\sum_{ \substack{ c \ge 0,   \\ (c,d)=1}} {\cal{D}} \mc{A} \right) \left(\sum_{ \substack{ c \ge 0,   \\ (c,d)=1}} {\cal{D}} \bar{\mc{A}} \right) ,
\eea
as a natural measure for 3D pure gravity path-integral. 

Fourth, we supersymmetrize $\mc{A}$. 
By introducing auxiliary fields and constructing 3d $\mc{N}=2$ vector multiplet $V = (\mc{A} , \sigma, D , \Lbar{\lambda} , \lambda)$, 
we supersymmetrize the Chern-Simons action
\bea
S_{SCS}[V]
=
S_{CS}[\mc{A}]
+ \int d^3 x \sqrt{g} \ \text{Tr} \Big( - \Lbar{\lambda} \lambda + 2 D \sigma \Big).
\label{scs}
\eea
Superficially, the additional terms in \eqref{scs} give no kinetic terms to the fields.
Therefore, we expect that this deformation does not break the relationship in \eqref{key}, and 
$\int {\cal{D}}\mc{A} \, e^{- S_{CS}[\mc{A}]} \approx \int {\cal{D}}V \, e^{- S_{SCS}[V]}$
holds. 
However there is a subtle issue; we will see later that the ``mass term'' $\sim \Lbar{\lambda} \lambda$ vanishes 
at the boundary from the supersymmetric boundary condition and as a result, there is a boundary localized 
decoupled fermion, at least, in the perturbative picture. 
Even though these additional degrees of freedom might bother us, 
there is still a big merit for the supersymmetrization, it enables us to evaluate the 
path-integral exactly by using the localization technique. 
Since the Chern-Simons theory is topological and has no local degrees of freedom, introducing 
auxiliary fermions are consistent with no essential modification assumption for the path-integral, 
aside from a renormalization of the coupling $k$ and introduction of additional boundary localized degrees of freedom. 

All of these give sequence of transformation as  
\bea 
& Z_{gravity}& = \int {\cal{D}} g_{\mu\nu} e^{- S_{gravity}} \to 
\sum \int {\cal{D}}g^{sector}_{\mu\nu} e^{- S_{gravity}} \nonumber \\
\label{sequence}
&\to & \int {(\sum \cal{D}}\mc{A} ) (\sum {\cal{D}}\mc{\bar{A}} ) e^{- \frac{i k}{4 \pi} S_{CS}[\mc{A}] + \frac{i k}{4 \pi} S_{CS}[\bar{\mc{A}}]}   \nonumber \\ 
&\to &  \int \left( {\sum \cal{D}}V \right) \left( {\sum \cal{D}} \bar{V} \right) e^{- \frac{i k}{4 \pi} S_{SCS}[{V}] + \frac{i k}{4 \pi} S_{SCS}[\bar{{V}}]    } \,, \nonumber   
\eea 
%
and we evaluate the path-integral exactly for the final expression 
by using the 
localization technique for the Chern-Simons theory with a boundary.  
If our assumptions are correct, we can evaluate the exact partition function for the pure gravity 
from the Chern-Simons theory.

Finally, we apply the localization principle to the calculation of 
\bea
\label{pathintegralgoal}
\sum_{ \substack{ c \ge 0,   \\ (c,d)=1}}
%
\int \mc{D}V e^{- \frac{i k}{4 \pi} S_{SCS}[{V}]} 
\equiv \sum_{ \substack{ c \ge 0,   \\ (c,d)=1}} Z_{c, d}
\label{ours}
\eea
to obtain the quantum gravity partition function (\ref{GRpathintegral}). 
In order to evaluate (\ref{pathintegralgoal}), we need to specify the boundary condition for the 
vector multiplet.  
Since 
the boundary condition in gravity side is asymptotic AdS metric, 
the vector multiplet must satisfy the Dirichlet boundary condition parametrised by $\tau$. 
We will soon write down the explicit formula for the boundary condition.  
Fortunately, explicit evaluation of the path-integral in (\ref{pathintegralgoal}) is conducted 
in \cite{Sugishita:2013jca} in the case that $\tau$ is purely imaginary \footnote{For path-integral under different boundary condition which is not Dirichlet, see \cite{Yoshida:2014ssa}.}. Therefore, we use that 
results and then we take the summation over $c$ and  $d$.

\section{Localization}
Given the argument that only one topology contributes is $D^2 \times S^1$, 
we briefly review the argument in \cite{Sugishita:2013jca} with the following metric on $D^2 \times S^1$: 
$ds^2 = d\theta^2 + \cos^2 \theta \left( d \varphi^2 + \tan^2 \theta d t_E^2 \right)$, 
for 
$0\leq\theta \leq \theta_0 < \pi/2\,,0\leq\varphi\leq 2\pi\,,0\leq t_E\leq 2\pi$, where 
$\theta = \theta_0$ is a boundary torus with purely imaginary 
$\tau = i \beta = i \tan \theta_0$ \footnote{We rename $\chi$ in \cite{Sugishita:2013jca} as $t_E$ for convenience.}.
For our interest, it is sufficient to state the details on vector multiplet $V$.
The vector multiplet on $D^2 \times S^1$ is constructed by a gauge field $\mc{A}_\mu$, a scalar $\sigma$, an auxiliary scalar $D$, two gauginos $\Lbar{\lambda}, \lambda$.
We can take the following Dirichlet boundary conditions
\bea
&& \mc{A}_\varphi \to a_\varphi \,,
\quad
\mc{A}_{t_E} \to a_{t_E} \,, \nonumber \\
\quad
&& \sigma \to 
0
\,, 
\quad  
\lambda \to
e^{-i (\varphi - t_E) } \gamma^\theta \Lbar{\lambda}\,, \,\quad \,\,\,\,\,
\label{bcb}
\eea
where all of them are proportional to $\sigma_3$, the Cartan generator for $SU(2)$.
The remaining component for the gauge field $\mc{A}_\theta$ is not fixed in their context. 
The fermion boundary condition $ \lambda = e^{-i (\varphi - t_E) } \gamma^\theta \Lbar{\lambda}$
kills the ``mass term'' $\sim \Lbar{\lambda} \lambda$ at the boundary. 
To understand this point in more detail, 
using the doubling trick, let us define the fermion fields $\psi_1$ and $\psi_2$ as 
$\psi_1(x) = \lambda_1 (x) \theta(x) + e^{-i (\varphi - t_E) }   \bar{\lambda}_1 (-x)  \theta(-x)$, 
$\psi_2(x) =  \bar{\lambda}_2  (x)  \theta(x) - e^{ + i (\varphi - t_E) }  \lambda_2 (-x) \theta(-x)$, 
such that these reflect the boundary condition at $x=0$, where 
$x \equiv \theta_0 - \theta$.  
Then 
the fermion ``mass term'' becomes 
$\Lbar{\lambda} (x) \lambda (x) 
= \mbox{sign}(x) \psi_1 (x) \psi_2 (x)$. 
This shows that only at the boundary, $x =  \theta_0 - \theta = 0$, 
this mass term vanishes. This is the typical ``domain wall fermion'' 
behaviour, where fermion are sharply localized at the boundary. 
Note that all of these arguments are under the $k \to \infty$ assumption, since 
we are discussing the classical ``mass term''.

For localization, we add super Yang-Mills action and it gives the localization 
locus $\mc{F} = 0$, where $\mc{F}$ is the field strength.  
Then, the classical contribution \cite{Banados:2012ue}, and the one loop determinant at the localization locus give the following formula \cite{Sugishita:2013jca}
\bea
  Z_{(c,d)}  &=& \int_{ \text{b.c.} \eqref{bcb}}
\mc{D} V
e^{- \frac{ik}{4 \pi} S_{SCS}[V]}  \notag \\
 &=& Z_{classical} \times Z_{one-loop} \,,\label{pert}
 \\
 Z_{classical}
 &=&
e^{  i k  \pi \tr (a_\varphi a_{t_E}) } \,, \notag 
 \\
Z_{one-loop} &=& 
 \prod_{m \in \mb{Z}} \Big(  m - \alpha(a_\varphi) \Big) 
 =
e^{ i \pi \alpha \left(a_\varphi  \right) }
-
e^{- i \pi \alpha \left(a_\varphi  \right)}, 
 \notag 
\eea
here we used the zeta function regularization for the infinite product (see the Appendix for the detail).  
This is the result of path-integral given the choice of 
$SL(2 ,\mb{Z} )$ coset parameters, $c$ and $d$.

\section{Boundary condition}
What we have to conduct next is to choose the boundary condition for $a_\varphi$ and  $a_{t_E}$ given the 
choice of 
$c$ and $d$, and $\tau$, then we sum over $c$ and $d$. 
However, for our purpose, 
it is enough to evaluate (\ref{pert})     
for the case of non-rotating BTZ black hole, {\it i.e.,}
$c =1$, $d=0$ choice. 
Then in the obtained expression, we replace 
\bea
- \frac{1}{\tau}
\to
\frac{a \tau + b}{c \tau +d} \,,
\quad
\mbox{with} \quad ad - bc =1 \,,
\label{SL2Ztr}
\eea
and conduct explicit summation over $c$ and $d$.

For the case of a BTZ black hole, we can find the boundary condition in \cite{Ammon:2012wc}.
By considering the BTZ black hole solution in the context of the Chern-Simons formulation,
the resulting gauge field satisfies
\bea
e^{2 \pi   \beta a_{\varphi}} = 
\begin{pmatrix}
-1 & 0 \\
0 & -1
\end{pmatrix}
\eea
for the non-rotating BTZ.
Then we can choose
\bea
a_\varphi
=
\frac{1}{2 i \beta} \sigma_3 = \frac{1}{2 \tau} \sigma_3 \,, \quad a_{t_E} = \frac{1}{2} \sigma_3 \,, 
\eea
%
%
where $\beta$ is the corresponding black hole inverse temperature, and we define the modulus $\tau = i \beta$ for later use.
By substituting this value, we arrive at the partition function around the non-rotating BTZ black hole solution as
\bea
Z_{(c=1, d=0)}
=
e^{\frac{1}{4} (k+ 2) \frac{2 \pi i}{ \tau}}
-
e^{\frac{1}{4} (k  - 2 ) \frac{2  \pi i}{ \tau} }
\label{BTZresult}
\eea

\section{Rademacher sum}
In order to conduct sum over $c$ and $d$, 
we conduct the replacement (\ref{SL2Ztr}) for the obtained expression (\ref{BTZresult}).  
Then we obtain for generic $c$ and $d$ as 
\bea
Z_{(c, d)} =
\label{cdpartitionfunction}
e^{ - {2 \pi i} \frac{1}{4}(k+2) \frac{a \tau + b}{c \tau +d}}
-
e^{ -{2 \pi i} \frac{1}{4} (k  - 2 ) \frac{a \tau + b}{c \tau +d} } \,. 
\eea
For the summation over $c$ and $d$, 
there is a good way to perform such summation with appropriate regularization, called \textit{Rademacher sums} \cite{rademacher:1939, Duncan:2009sq}.
After taking such regularization, we arrive at the following ``holomorphic partition function for the pure-AdS gravity"
\bea
Z_{hol} [q] &\equiv& Z_{(0, 1)} (\tau) + 
\sum_{ \substack{ c > 0,   \\ (c,d)=1}}
\Big(
Z_{c, d}
(\tau)
-
Z_{c, d}
( \infty )
\Big)
 \nonumber \\
&=&
R^{\left(- {k_{eff}}/{4} \right)} (q)
-
R^{\left(-k_{eff}/4 +1 \right)} (q) \,, 
\label{finalZ}
\eea
where 
$k_{eff} \equiv  k + 2 $, and 
\bea
R^{(m)} (q) &\equiv& e^{2 \pi i m \tau} + \sum_{ \substack{ c > 0,   \\ (c,d)=1}}
\left( e^{2 \pi i m \frac{a \tau + b}{ c \tau + d}} -   e^{2 \pi i m \frac{a}{ c}}\right) \nonumber \\
&=& q^m + (\mbox{const.}) + \sum_{n=1}^\infty c(m,n) q^n \,,\\
c(m,n) &\equiv& \sum_{ \substack{ c > 0, \\  (c,d)=1, \\ d \, mod \, c}} e^{2 \pi i (m \frac{a}{c} + n \frac{d}{c} )} 
\sum_{\nu=0}^\infty \frac{\Big( \frac{2 \pi}{c} \Big)^{2\nu+2}}{\nu ! (\nu+1)!} (-m)^{\nu +1}n^\nu  \,, \nonumber 
\eea
and $q \equiv e^{2 \pi i \tau}$. 
Subtraction by $Z_{c, d}( \infty )$ is for regularization. 
For $m=-1$ case, $(\mbox{const.})= 12$ \cite{rademacher:1939}. 
This is our main result.  
%
%
%
%
Note that in case $m$ is not an integer, $q^m$ picks up a phase under the $\tau \to \tau +1$. 
Since one can always tune a moduli $\tau$ so that $q^m$ becomes a dominant term in $R^{(m)}(q)$ for $m<0$, 
$m$ needs to be an integer, otherwise it contradicts with the 
modular invariance. 
In fact, this function is well studied in \cite{Duncan:2009sq} and it is shown there that 
under the assumption that $R^{(m)} (q)$ converges, $R^{(m)} (q) $ vanishes for fractional number $m = -g/h$, 
with integers $g$, $h$, unless $h=1$. 

$m=-1$ is of special interest, and $R^{(-1)}(q)$ is Klein's $J$-function; $J(q)$.

\section{Quantum Gravity Partition Function}
%
Let us discuss our final expressions (\ref{finalZ}): 
%
%
The mathematical fact that $m$ need to be integer under the assumption of convergence and modular 
invariant is interesting since it is exactly the quantisation of $k_{eff}/4$. 
On the other hand, in the Chern-Simons theory, we take the quantization condition for $k$ as $k/4 \in \mathbb{Z}$. 
This is interesting since except for the difference between $k$ and $k_{eff}$, they are the same condition! Therefore, 
we expect that the relationship $c = 6 k_{eff}$ to hold. 

Furthermore, for $k_{eff}/4=1$ case, in which we expect $c=24$ \footnote{We expect that 
due to the auxiliary fields we introduced, the new relationship $c = 6 k_{eff}$ to hold.}, 
using our sequence 
and relation (\ref{pathintegralgoal}), the final expression simply becomes 
\bea
Z_{gravity} = J(q) J(\bar{q})
\eea
\textit{up to constant number shift for $J(q)$ and $J(\bar{q})$.}
This partition function is exactly the one  
predicted by Witten in \cite{Witten:2007kt}! 
Since the term $q^{-1}$ represents the AdS vacuum, the constant term $\sim q^0$ 
comes from the level 1 descendant of AdS vacuum. BTZ black holes cannot contribute to this constant term 
since they exist regularly only $M > 0$, $J \ge 0$.  
If dual boundary theory exists, 
this suggests that the constant term in partition function should vanish, 
since Virasoro operator $L_{-1}$ acting on vacuum vanishes. 
%
On the other hand, in our direct path-integral, there is a priori no principle to determine this constant 
since it depends on how we regularize the $c$, $d$ summation. We leave this as a open question. 

For integer $k_{eff}/4$ with $k_{eff}/4 >1$ case,   we face negative coefficients in the $q$ expansion. 
For example, for $k_{eff}/4 = 2$, we obtain 
\bea
\label{c48case}
Z_{hol}[q] = q^{-2} - q^{-1} + \sum_{n=0}^\infty c_{2, n} \, q^n \,. 
\eea
where $c_{2, n}$ is some constant. 
To interpret this negative coefficients, we propose two possibilities. 1. Dual CFT does not exits. 
2. Dual CFT exists, though it possess fermion degrees of freedom.  
We believe that if dual CFT exists, then it 
should possess fermion degrees of freedom. 
As we have discussed, our theory possess boundary decoupled fermion at least perturbatively.  
This is reflected in our results;  for example in the $AdS_3$ background ($a = d = 1$, $b = c = 0$) our result (\ref{cdpartitionfunction}) can be written as 
$Z_{(c=0, d=1)}  = Z_{B-fermion}  \times
q^{- k_{eff}/4} \, \Pi_{n=2}^\infty (1 - q^n)^{-1} $
where $Z_{B-fermion} \equiv \Pi_{l=1}^\infty \left( 1 - q^l \right)$ is the partition function 
of the boundary localized free fermion, whose existence we have discussed before.  
Note that contribution $q^{- k_{eff}/4}  \Pi_{n=2}^\infty (1 - q^n)^{-1}$ is exactly the 
contribution from classical $AdS_3$ saddle point and its Virasoro descendant \cite{Maloney:2007ud, Giombi:2008vd}. 

\section{Summary and Discussion}
Let us close this paper
with discussion. 
We applied the recently developed localization technique to the 3D gravity 
and obtained the remarkable $J$-function, predicted by Witten. 
Big assumption was made that quantum pure gravity in AdS$_3$ can be formulated non-perturbatively 
by summing over boundary conditions for Chern-Simons theories. 
But given that, localization gives a nice justification for our procedure: 
First, only the localization locus $\mc{F} = 0$ contributes in the path integral, but 
$\mc{F} = 0$ is the Einstein's equation, and the only topology allowed there is a solid torus.  
Furthermore, the Rademacher sum arises naturally because 
we have to sum over all the field configurations satisfying localization locus $\mc{F} = 0$. 
Second, our calculation for the path integral is exact, therefore 
does not rely on the perturbative method. 
Third, gravity action in terms of the Chern-Simons theory is decomposed as holomorphic and anti-holomorphic part, so 
holomorphic factorization is natural. These are the main reasons why our localization method evades previous 
obstacles and we succeed in obtaining the $J$-function.   
However, clearly more studies are needed. 
We leave these as future problems.

\acknowledgments{%
The work of NI was supported in part by 
JSPS KAKENHI Grant Number 25800143.  
The work of AT was supported in part by the RIKEN iTHES Project.}

\,

\appendix

\section{Derivation}
%
Here, we derive the results for $Z_{one-loop}$ 
in \eqref{pert} in detail: 
%
\bea
Z_{one-loop} &=& \prod_{m \in \mb{Z}} \Big(  m - \alpha(a_\varphi) \Big) \nonumber \\
&=&
- 
\prod_{m =1}^\infty (-m^2)
\times
\alpha(a_\varphi)
 \prod_{m =1}^\infty \Big(
 1- \frac{[\alpha(a_\varphi )]^2}{m^2}
 \Big)
\notag \\
&=&
- 
\prod_{m =1}^\infty (-m^2)
\times
\frac{1}{\pi} \sin \pi \alpha(a_\varphi )  \,,
\label{A1eq}
\eea
then by using the zeta function regularization, 
\bea
&& - \prod_{m =1}^\infty (-m^2) 
 =  - \,  e^{ \sum_{m =1}^\infty  \left( \pm i \pi + 2  \log m\right) } \nonumber \\
& = & -  e^{ \pm i \pi \zeta(0) + 2 (- \zeta ' (0))} 
= \pm \, 2 \pi i \,, 
\eea 
we have 
\bea
Z_{one-loop}
&=& \pm \left(
e^{i \pi \alpha(a_\varphi )}
-
e^{-i \pi \alpha(a_\varphi )} \right).
\eea
There is an ambiguity for the overall sign, which is undetermined. 
However this overall ambiguous sign is independent on the choice of 
$c$ and $d$, the cosets of $SL(2 ,\mb{Z} )$ parameter. It gives at most 
overall sign ambiguity of the final path-integral expression. 
Therefore 
in this paper, we neglect this sign.  
\\


\end{document}